\documentclass[conference]{IEEEtran}
\IEEEoverridecommandlockouts
\usepackage{amsmath}
\usepackage{amsmath,amssymb,amsfonts}
\usepackage{graphicx}
\usepackage{textcomp}
\usepackage{xcolor}
\usepackage{enumitem}
\usepackage{algorithm}
\usepackage{algpseudocode}
\usepackage{caption}
\def\BibTeX{{\rm B\kern-.05em{\sc i\kern-.025em b}\kern-.08em
    T\kern-.1667em\lower.7ex\hbox{E}\kern-.125emX}}

\begin{document}

\title{Joint Route Selection and Power Allocation in Multi-hop Cache-enabled Networks\\

\thanks{This work was supported by Grant No. SGS23/171/OHK3/3T/13.}
}

\author{\IEEEauthorblockN{Emre Gures,
		Pavel Mach \textsuperscript}
	\IEEEauthorblockA{Faculty of Electrical Engineering, Czech Technical University in Prague, Prague, Czech Republic}
	   \IEEEauthorblockA{\{guresemr, machp2\}@fel.cvut.cz}
	}%
\maketitle

\begin{abstract}
The caching paradigm has been introduced to alleviate backhaul traffic load and to reduce latencies due to massive never ending increase in data traffic. To fully exploit the benefits offered by caching, unmanned aerial vehicles (UAVs) and device-to-device (D2D) communication can be further utilized. In contrast to prior works, that strictly limits the content delivery routes up to two hops, we explore a multi-hop communications scenario, where the UAVs, the UEs, or both can relay the content to individual users. In this context, we formulate the problem for joint route selection and power allocation to minimize the overall system content delivery duration. First, motivated by the limitations of existing works, we consider the case where the nodes may transmit content simultaneously rather than sequentially and propose simple yet effective approach to allocate the transmission power. Second, we design a low-complexity greedy algorithm jointly handling route selection and power allocation. The simulation results demonstrate that the proposed greedy algorithm outperforms the benchmark algorithm by up to 56.98\% in terms of content delivery duration while it achieves close-to-optimal performance. 
\end{abstract}


\section{Introduction}
The emerging 6G-based mobile networks will have to cope with unprecedented data transmissions originated form plethora of communicating devices, such as smartphones, sensors, vehicles, or any internet of things (IoT) devices. This will inevitably pose high requirements on provided data rates over backhaul and experienced latencies. To alleviate backhaul load and to enable low latencies, caching of popular content seems to be a very promising approach \cite{zyrianoff2022iot}. Obviously, the popular content should be cached in proximity of a user equipment (UE), such as at a ground base station (GBS).

To improve systems-wide performance, the unmanned aerial vehicles (UAVs) can be exploited as caching servers as well \cite{duong2022uav}. Such cache-enabled UAVs can store popular contents, thus reducing content delivery duration and backhaul traffic load \cite{zhao2019caching}. The UAV caching can be particularly beneficial during peak hours to offload traffic of the GBSs or to mitigate severe shadowing in urban or mountainous scenarios by leveraging their ability to establish line-of-sight (LoS) connections with ground nodes. The mobile networks can also benefit from the device-to-device (D2D) functionality of UE; to transmit data to other relaying UEs (RUEs) \cite{mach2022device}. In such cached-enable, UAV-assisted, and D2D-enabled networks, the main challenges are the selection of proper route over which the content should be traversed in order to reach the UEs, power allocation, content placement, or UAVs' deployment. 

The problem of route selection and power allocation in cache-enabled UAV-assisted D2D-enabled cellular network is considered in many works targeting various objectives, including optimization of minimum secrecy rate among requesting UEs \cite{ji2020joint}, sum throughput \cite{al2021throughput}, and energy efficiency \cite{qi2022joint}. Whereas the afore-mentioned studies \cite{ji2020joint,al2021throughput} confine their scope to direct communication, \cite{qi2022joint} enables two-hop communication using UAV-relaying. Still, as the distance between the source and target nodes increases or the communication environment deteriorates, direct communication or even two-hop communication with single UAV relay is generally insufficient to reduce the content delivery duration \cite{zhang2018trajectory}. For instance, in a densely populated urban environment, wireless communication links are susceptible to blockage by tall buildings. Consequently, the mitigation of such link blockage problems typically necessitates the employment of multi-hop communication incorporating both UAV relays and the RUEs in order to provide sufficient degrees of freedom. 

In addition, some studies propose different methods to minimize content delivery duration. For example, in \cite{wang2021deep}, a deep deterministic policy gradient-based caching placement strategy is proposed. In \cite{luo2023cost}, the UAV deployment and content placement are jointly studied. However, in both \cite{wang2021deep} and \cite{luo2023cost}, transmission nodes send contents sequentially rather than simultaneously. Nevertheless, this approach does not minimize transmission duration, as the duration is not linearly proportional to the allocated transmission power.

In this article, we aim to cover the gaps of the existing related works. We formulate the problem as joint route selection and power allocation problem minimizing sum content delivery duration. Unlike \cite{ji2020joint,al2021throughput,qi2022joint}, where up to 2-hop communication is enabled, we target multi-hop scenario (i.e., more than 2 hops). To the best of our knowledge, no multi-hop transmission route selection for delivery of cached content has not been considered so far. The content delivery over multi-hop transmission is expected to decrease latencies in urban scenario with many potential obstacles in the communication path. Note that a deployment of many conventional fixed GBSs can also decrease latencies, but at significant installment cost while these GBS would still not be able to cope with dynamic requirements of users. In addition, compared to \cite{wang2021deep} and \cite{luo2023cost}, we assume that each transmitting node can send several contents simultaneously to further decrease delivery of cached content. 

The enabling of multi-hop communication, however, makes also the selection of the optimal transmission routes over which content traverse to the requesting UE a very challenging task. The reason is that there are many potential transmission routes for delivering the content to individual UEs. The route selection also affects the power allocation at GBS, UAVs, and/or RUEs that can be potentially involved in delivery of multiple contents, thus impacting the overall content delivery duration. As a result, the route selection and power allocation must be handled jointly. We summarize our main contributions to address above-listed challenges  as follows:
\begin{itemize}[leftmargin=0.4cm]
	\item We propose the transmission power allocation managing the splitting of transmission power budget by each transmitting node (i.e., the GBS, the UAV, or RUE) to each content currently being sent. We guarantee the continuous utilization of the entire transmission power by the transmitting nodes, resulting in a reduction in the overall transmission duration.
	\item We propose a low-complexity greedy algorithm that jointly considers the route selection while exploiting proposed power allocation.
	\item We show the proposed greedy algorithm decreases overall content delivery duration by up to  56.98\% in comparison to the benchmark algorithm. At the same time, we show that greedy algorithm yields close-to-optimal performance. 
\end{itemize}
The rest of the paper is organized as follows. Section II introduces the system model and Section III formulates the problem. In Section IV, we present the proposed greedy based joint route selection and power allocation algorithm. Section V introduces the simulation model and competitive algorithms. In Section VI, simulation findings are examined and discussed. Finally, we briefly summarize the findings in Section VII.

\section{System Model}

This section describes the network model, cache placement model, and provides details about content delivery duration.

\subsection{Network Model}
\begin{figure}
	\centering
	\includegraphics[scale=0.15]{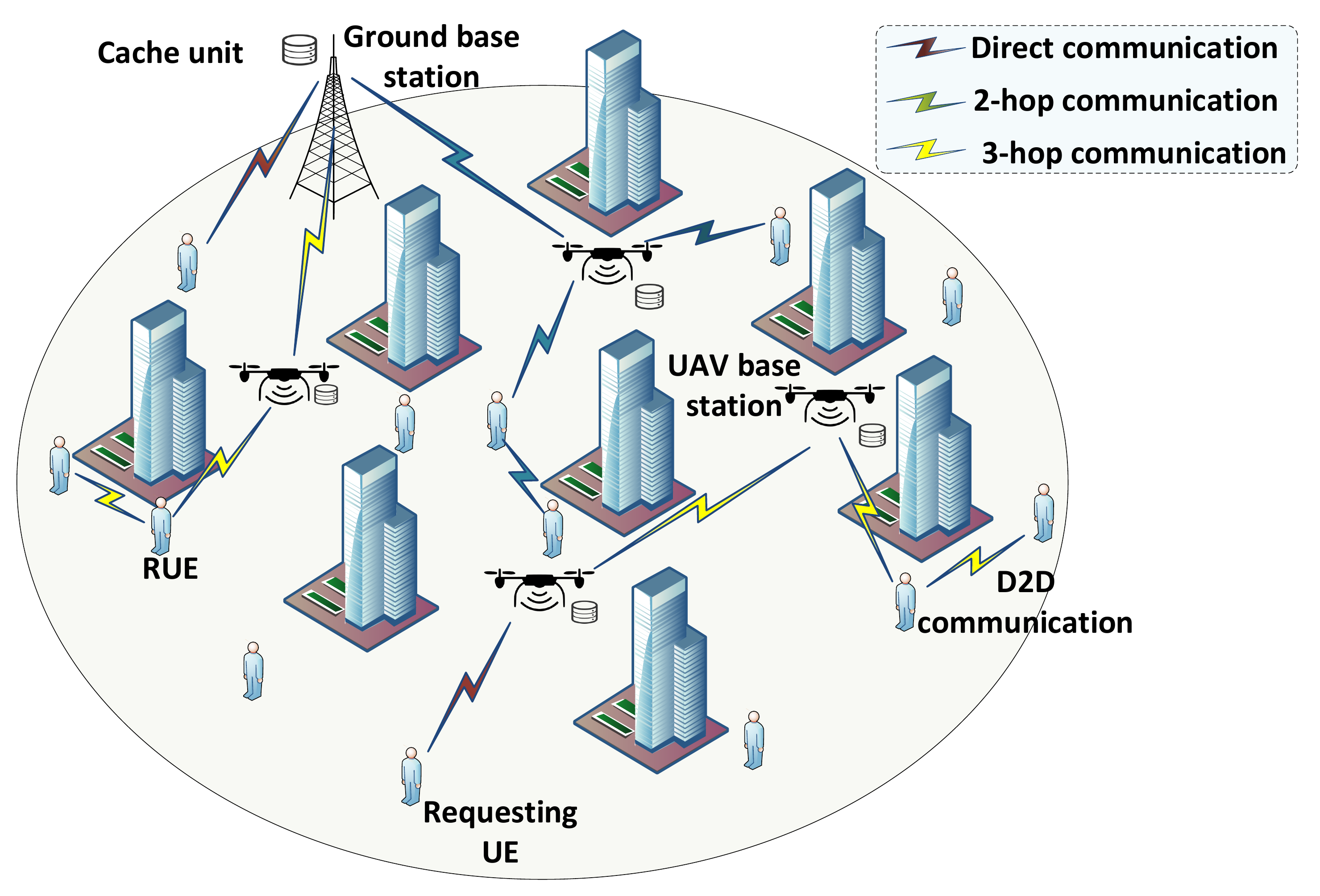}
	\caption{A multi-hop cache-enabled network in urban area.}
	\label{fig:network_multihop1}
\end{figure}
We consider a cache-enabled, UAV-assisted, and D2D-enabled cellular network, as depicted in Fig. \ref{fig:network_multihop1}. The network is composed from $K+1$ BSs including one GBS and $K$ UAVs acting as flying BSs. Further, we assume there are $U$ UEs in the system from which $N$ are asking for delivery of a content. We consider that the UEs currently not requesting the content can serve as the RUEs, i.e., there are $R=U-N$ RUEs. Note that relaying accomplished by the UEs can be facilitated by D2D relaying concept \cite{mach2022device}. The content can be, then, delivered to the UEs: \emph{i}) directly by the GBS, \emph{ii}) directly by the one of the UAVs that can reduce content delivery duration by caching popular content requested by the UEs in cache repositories, or \emph{iii}) by relaying the content either via the UAV and/or RUEs. 

Both the UAVs and RUEs operate in half-duplex mode \cite{mach2021power}. We limit the number of relaying nodes for delivery of content up to two (i.e., up to 3-hop communication) as more relaying nodes usually yield only minimal gains in delivery duration while the complexity is increased significantly \cite{mach2022device}. Considering that the content can be delivered directly by the GBS or UAVs, and through 2-hop or 3-hop routes, there is $M$  possible transmission route options for any requesting UE. 

\subsection{Cache Placement Model}
We assume the UEs can pick out of $F$ contents, where the size of \emph{f}-th content is $S_{f}$. The users’ requests are determined by a content popularity distribution, denoted as $p=\{p_{1},p_{2},…,p_{F}\}$. Here, $p_{f}$ represents the probability of a request for content \emph{f} whilst $\sum_{f} p_{f}=1 $. The popularity of the contents is inherent to them, and it is assumed that the distribution of their popularity follows the Zipf distribution \cite{zhang2023joint}, described as ${p_{f}} = f^{-\gamma}/\sum_{j=1}^F j^{-\gamma}$,
where $\gamma$ is the Zipf exponent indicating the degree of skewness in popularity.

 We assume all the contents are stored at GBS while the UAVs store only a subset of the content catalog due to their limited storage capacity. In other words, the UAVs can only cache a part of the files. In accordance with the caching probabilities $p_{f}$, each \emph{k}-th UAV stochastically generates a collection of contents $F_{k}$ to be cached, as outlined in the probabilistic content caching approach introduced in \cite{blaszczyszyn2015optimal}. 
\subsection{Content Delivery Duration}
The subset of nodes that will deliver content requested by the UE depends on both the cache status of the UAVs and the condition of the wireless channels. We assume that the entire bandwidth $B$ is divided into $N$ orthogonal channels, so that the \emph{n}-th UE requesting a content is assigned a channel with a bandwidth of $B_{n}$. Notice that the bandwidth allocation falls outside the purview of this article, and our proposed method is applicable to any arbitrary allocation of bandwidth. Hence, we assume that $B$ is split equally among $N$ requesting UEs. 

The content delivery duration for the \emph{f}-th content requested by the \emph{n}-th UE whilst using the \emph{m}-th transmission route including $H_{m}$ hops is defined as follows:
\begin{equation}
	{t_{n,f,m}} = \sum_{h_{m}=1}^{H_{m}} \frac{S_{f}}{B_{n} \log_2 (1+ \frac{p_{f}^{h_{m}} g^{h_{m}} } {B_{n}(\sigma_{0}+I_{b})} )},
\end{equation}
where $p_{f}^{h_{m}}$ denotes the transmission power of the node exploited in the $h_{m}$-th hop on the \emph{m}-th transmission route to deliver the \emph{f}-th content to the requesting UE, $g^{h_{m}}$  represents the channel gain at the $h_{m}$-th hop of the \emph{m}-th transmission route, $\sigma_{0}$ is the noise power, and $I_{b}$ represents the background interference received from neighboring transmitters.

We assume that each transmitter (GBS, UAV, or RUE) can transmit multiple contents simultaneously. Note that this is reasonable assumption as the GBS usually transmits different data to different users albeit at different frequencies. In such case, the delivery duration in (1) is affected by the number of contents sent at the same time since available transmission power budget at the transmitter has to be distributed among all contents. To ensure that maximum transmission power is not violated, we introduce two indicators $y_{k,f,m}$ and $v_{r,f,m}$, to keep track on whether the \emph{k}-th BS and/or the \emph{r}-th RUE are in the \emph{m}-th transmission route selected for the \emph{f}-th content delivery, respectively. The $y_{k,f,m}$ and $v_{r,f,m}$ are equal to 1 if the \emph{k}-th BS and the \emph{r}-th RUE are on the \emph{m}-th transmission route, respectively, and 0 otherwise.

\section{Problem Formulation}
The objective of this article is to jointly optimize route selection \emph{X$^*$} and power allocation \emph{P$^*$}, aiming to minimize the content delivery duration of the requesting UEs. We denote the route selection indicator by $x_{n,f,m}\in \{0,1\}$, where $x_{n,f,m}=1$ when the \emph{n}-th UE requests the \emph{f}-th content from the \emph{m}-th route, and $x_{n,f,m}=0$ otherwise. Then, the targeted optimization problem is formulated as follows:
	\begin{align}
		X^*, P^* = \min_{X,P} \sum_{n}\sum_{f}\sum_{m} x_{n,f,m}t_{n,f,m}                                     \\
		\textrm{s.t.} \quad \text{(c1)} \quad \sum_{n}\sum_{f} P_{k}	y_{k,f,m}\leq P_{k}^{max}, \forall k     \\
		\textrm{    }  \quad  \text{(c2)} \quad  \sum_{n}\sum_{f} P_{r}   v_{r,f,m}\leq P_{r}^{max}, \forall r   \\
		\textrm{    } \quad \quad \text{(c3)} \quad  \sum_{m} x_{n,f,m} \leq 1, \forall n,f \quad \quad \quad \ \\
		\textrm{    } \quad \quad \quad \text{(c4)} \quad H_{m} \leq H_{max}, \forall m \quad \quad \quad \quad \ \ \
	\end{align}
where (c1) and (c2) ensure that the total transmission power allocated by any BS and RUEs sending the content(s) does not exceed $P_{k}^{max}$ and $P_{r}^{max}$, respectively, (c3) ensures each content traverse via one unique route, and (c4) limits the maximum number of hops $H^{max}$ for any \emph{m}-th route.

The optimization problem can be classified as a mixed integer non-linear programming (MINLP) problem due to integer constraint (c3) while the objective function is non-linear with respect to power allocation (see expression of $t_{n,f,m}$ in (1)). These problems are NP-hard, meaning that there is no known polynomial-time algorithm for solving them optimally. The problem is very complicated also since route selection \emph{X} and transmission power allocation \emph{P} are coupled problems that should be solved jointly, as the selection of transmission routes depends on the relationship between the requesting user and the nodes, while power allocation affects this relationship. 

One way to find the optimal solution is to use a brute-force search. The brute-force has, however, huge complexity ($O(M^N)$) even for small number of UAVs, the UEs requesting content and the RUEs since all possible route combinations have to be tested. One approach to reduce the complexity of brute-force search is to exclude ``bad" combinations that are not able to outperform direct route, e.g., the combinations where the content delivery duration over any hop is actually higher than a direct route. Unfortunately, even this reduction does not make the problem solvable in polynomial time. 

\section{Proposed Scheme for Content Delivery}
This section explains the power allocation principle that is further exploited in proposed greedy algorithm.
\subsection{Power Allocation}
\begin{figure}[t!]
\centering
\includegraphics[scale=0.55]{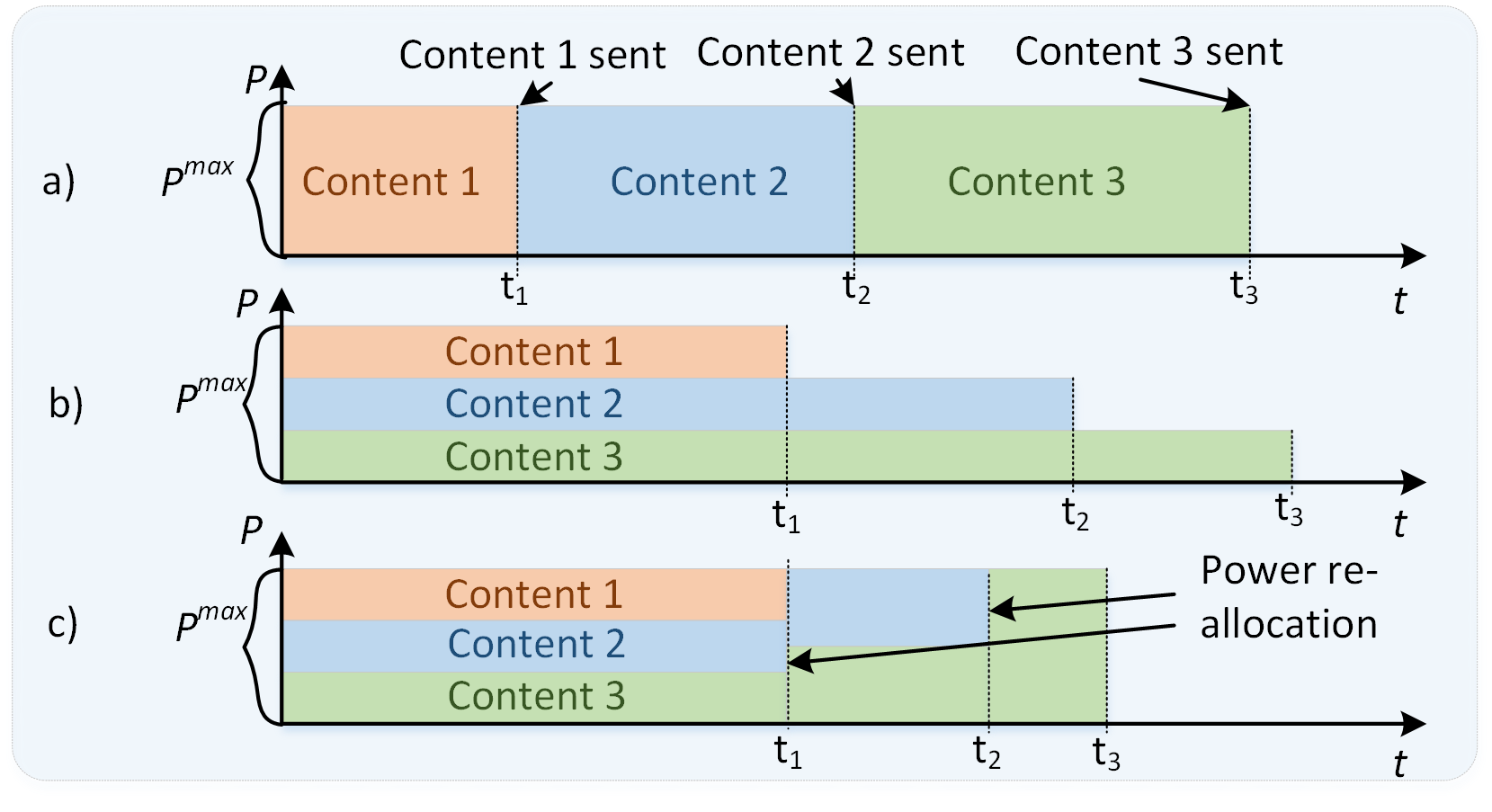}
\caption{Principle of power splitting among multiple contents. } 
\label{fig:Drawing2}
\end{figure}
Let's first explain the proposed power allocation principle adopted during the following route selection process. The proposed power allocation deals with the problem when any transmitting node is about to transmit more than one content simultaneously. Of course, each content can be simply sent in a sequential manner, where the transmitting node allocates whole power budget to transmit the first content, then to the second content, and so on (see Fig. \ref{fig:Drawing2}a), where content 1 is send first and then follow content 2 and content 3) \cite{luo2023cost}. This option is not minimizing transmission duration since the duration is not linearly proportional to the transmission power allocated for the transmission (see (1)). 

To this end, we propose an approach where each transmitting node can, in fact, transmit several contents simultaneously in order to decrease overall transmission duration of the contents while constraints (c1) and (c2) in (2) are not violated. We assume here that the whole transmission power is split equally among individual contents that are being currently transmitted. Note that the optimization of the power splitting itself is left for future research due to limited page number. Since, in general, each content can have different size and can be transmitted over different channels, the transmission of each content can take different transmission duration as well. If the transmission power allocated to each content would remain the same, however, this would inevitably prolong the transmission duration of some contents, as transmission power used to send each content is not always fully used, as illustrated in Fig. \ref{fig:Drawing2}b. 

Hence, we propose that the transmission power allocated to each content is redistributed among the contents after each individual content is fully transmitted to the receiving node. In other words, we ensure that the whole transmission power is used all the time by the transmitting node and, thus, the total transmission duration is reduced by our approach. This is illustrated in Fig. \ref{fig:Drawing2}c where after content 1 is already transmitted, power is re-allocated to transmit content 2 and 3. Then, when also content 2 is finished, the whole transmission power can be allocated just to transmit the rest of content 3.

In the sequel, we describe in detail the route selection process, where the above-explained power allocation is exploited.

\subsection{Proposed Joint Route Selection and Power Allocation}
The main challenge of the problem in (2) is that there are plenty of possible routes over which each content can be sent to the individual requesting UEs. Besides, the selection of route is affected by the number of contents sent by each node and, subsequently, the content delivery duration. The reason is that if any node is sending multiple contents simultaneously, it has inevitably less transmission power that can be allocated to each content, which affects the selection process. Consequently, the route selection and power allocation at individual nodes involved in content delivery must be handled jointly. Thus, we propose a greedy approach solving the problem at reasonable complexity, summarized in Algorithm 1. 

At the very beginning, Algorithm 1 selects for each requesting UE the potential source options ($L_{n}$) distinguishing which nodes have the requested content (see line 1 in Algorithm 1). Obviously, if no UAV has the requested content, the GBS is always selected as the source option. Otherwise, more than one source options are available. Then, for any \emph{n}-th UE requesting \emph{f}-th content, Algorithm 1 calculates content delivery durations for all source options, denoted as $t_{n,f,m'}$, where $m'$ indicates only direct routes out of all potential routes (line 2). Next, the direct route selection indicator $x_{n,f,m'}$ is initialized to 0 for all \emph{n} and $m'$, indicating that no UE requesting content \emph{f}-th has yet selected any direct route (line 3). The direct route for each requesting UE is selected in a sequential manner based on the content delivery duration. In particular, at each iteration, the direct route with the lowest content delivery duration, denoted as $t_{n,f,d}$, is selected for each UE (line 5) and $x_{n,f,m'}$ is set to 1 to indicate that the direct route is selected (line 6). To ensure that the selected route is not selected repeatedly, the content delivery durations of each direct transmission route for the \emph{n}-th requesting UE are set to infinity (line 7). Following the power re-allocation at the BS on the selected direct route according to the route selection (done in line with proposed re-allocation shown in Fig. \ref{fig:Drawing2}c), content delivery durations are updated for the UEs for which the direct route is not yet selected (line 8). The steps in lines 5-8 are repeated each requesting UE is assigned a direct route.

After the selection of direct route is finished, a decision if a multi-hop route would be more beneficial in terms of delivery content duration follows. Thus, we first calculate a multi-hop gain representing the duration reduction if any \emph{n}-th UE requesting the \emph{f}-th content would select any $m''$-th multi-hop route (i.e., $t_{n,f,m''}$) instead of a direct one as (line 10):
\begin{equation}
	G_{n,f,m''},=
	\begin{cases}
		t_{n,f,d}-t_{n,f,m''}-t_{m''}, & \text{if}\ G_{n,f,m''}>0\\
		0, & \text{otherwise}\\
	\end{cases}  
\end{equation}
where $t_{m''}$ represents an overall prolongation of content delivery duration for other UEs that are already using some nodes at $m''$-th route to deliver their content. The reason why $t_{m''}$ needs to be considered in (7) is that if some transmitting node(s) at $m''$-th route are already involved in transmission of other contents for different UEs, the transmitting node(s) must split transmission power as discussed in Section III.A, resulting in prolongation of content delivery duration. On the other hand, if transmitting node in $m''$ route send no content at the moment $t_{m''}$ is 0. If resulting $G_{n,f,m''}$ in (7) is positive, a multi-hop route is more beneficial to the requesting UE. Otherwise, direct route is kept and $G_{n,f,m''}$ is set to 0, since multi-hop route would increase the overall delivery duration. 

In the next step, the multi-hop gains between all requested UEs and transmission routes are inserted into a multi-hop gain matrix \emph{\textbf{G}}, which is of dimension $N\times M-(K+1)$ (excluding direct routes), as showed in line 11. In addition, route selection indicator $x_{n,f,m''}$ is initially set to 0 for all \emph{n} and $m''$ to indicate that no requesting UE has selected yet any multi-hop transmission route to receive the \emph{f}-th content (line 12). 

Now, the following steps are repeated as long as there is at least one positive entry in \emph{\textbf{G}}. The Algorithm 1 first selects the maximum multi-hop gain in \emph{\textbf{G}}, as this selection decreases the overall content delivery duration by the highest degree (i.e., Algorithm 1 finds max($G_{n,f,m''}$), line 14). Then, $x_{n,f,m''}$ is set to 1 to indicate direct route is changed to multi-hop route (line 15). Then, all entries in \emph{n}-th row of \emph{\textbf{G}} are set to 0, as this UE cannot select any other transmission route to receive the requested content (line 16). 

\begin{algorithm}[b]
\caption{\scriptsize Proposed Joint Route Selection and Power Allocation}
\scriptsize 
\begin{algorithmic} [1]
    \State Generate a list of potential source options $L_{n}$ for each UE \emph{n} 
    \State Calculate $t_{n,f,m'}$, $\forall n$, $\forall m'\in L_{n}  $ 
    \State Set  $x_{n,f,m'}=0$, $\forall n,m'$ 
    \Repeat 
    \State $t_{n,f,d}\leftarrow \min (t_{n,f,m'})$, $\forall n$
    \State $x_{n,f,m'}=1$  
    \State Set \emph{n}-th row in $t_{n,f,m'}$ to $\infty$ 
    \State Update $t_{n,f,m'}$ with power re-allocation 
    \Until all UEs requesting content are assinged to a direct route  
    \State Calculate $G_{n,f,m''}$, $\forall n,m''$ 
    \State Create matrix \textbf{\textit{G}} 
    \State Set $x_{n,f,m''}=0$, $\forall n,m''$   
    \While  {$\max(G_{n,f,m''})>0$}
    \State $\{n,f,m''\}\leftarrow \max(G_{n,f,m''})\in \textbf{\textit{G}}$ 
    \State $x_{n,f,m''}=1$  
    \State Set \emph{n}-th row in \textbf{\textit{G}} to 0   
    \State  Update \textbf{\textit{G}}, using (7) with power re-allocation      
    \EndWhile
\end{algorithmic}
\end{algorithm}

Next, the Algorithm 1 also has to update all remaining positive entries in \emph{\textbf{G}} (if any) that include nodes of $m''$-th multi-hop route according to (7) (line 17). The reason is that the potential gains are decreased as transmission power would need to be divided among more contents. Of course, this also means that the content delivery durations of the UEs that have already selected multi-hop route including these nodes in their transmission routes would be affected. Thus, the increase in overall content delivery duration should be less than the potential reduction achieved by new UEs selecting transmission routes. Therefore, this effect is reflected in the recalculation in \emph{\textbf{G}} via $t_{m''}$ with the updated power allocations.   

Algorithm 1 repeatedly executes lines 14-17 until there are no more positive inputs in the relaying gain matrix \emph{\textbf{G}}. In each iteration, the system updates the power allocation and selects the route with the highest relaying gain accordingly. Note that requesting UEs that do not select a multi-hop transmission route receive their content via the direct route.

Algorithm 1 has, in the worst case, a time complexity of $O([M-(K+1)]\sum_{n=1}^{N} n)= O(N^2 [M-(K+1)])$, which means that the time it takes to run the algorithm grows linearly with the number of transmission routes ($M$) and quadratically with the number of requesting UEs ($N$). 

\section{Simulation Setup and Competitive Algorithms}
To investigate and validate the proposed model, we performed simulations in MATLAB. We consider a $500\times500$ m reference cell with multiple buildings to imitate an urban environment (see \cite{mach2021power} for more details). The height of each building is randomly generated between 20 and 29 m. The GBS is placed in a fixed position in the upper left corner of the building, near the cell center. Furthermore, 4 UAVs are placed in the simulation area. The positions of the UAVs are determined by K-means clustering, based on the positions of the UEs. Last, we consider 100 UEs from whom up to 30 are requesting the content while other can serve as RUEs. Since the GBS and UAVs are above buildings, they communicate with each other via LoS. On the contrary, the communication path between the GBS/UAVs and the UEs can be obstructed by one or several buildings, each attenuating the signal by additional 20 dB. We run the simulations for 1000 drops. In each such drop, we randomly generate the building heights, the current contents cached at the UAVs, and the UEs locations according to which the UAVs locations are determined. Then, we average the simulation results over all drops. The simulation parameters are summarized in Table I.

The section with results presents the findings of two proposed approaches. The first approach performs joint route selection and power allocation while proposed power allocation is not adopted (labeled as ``Proposal: w/o PA"). The second approach is the whole proposal where also proposed power allocation is used (Proposal: with PA). The performance of the proposal is compared with benchmark algorithm (labeled as ``Benchmark"), where the transmission route is selected based on caching status and maximum biased received power \cite{gu2021association}. Further, we show the performance of brute-force algorithm to show how close to optimal the proposed greedy approach performs. The brute-force algorithm checks all possible transmission routes for each requesting UE and selects the transmission routes that provide the optimal average content delivery time. We illustrate the results for both the case w/o proposed power allocation (Brute-force: w/o PA) and with power allocation enabled (Brute-force: with PA). Note that the brute-force algorithms are only shown for a limited number of requesting UEs because the complexity of testing all transmission route combinations grows exponentially with the number of UEs. 
\begin{table}[]
	\centering
	\caption{Simulation parameters}
	\begin{tabular}{|l|l|}
		\hline
		\textbf{Parameter}                                                 & \textbf{Value}                                             \\ \hline
		Carrier frequency                                         & 2 GHz                                             \\ \hline
		Simulation area of the reference cell                     & 500x500 m                                         \\ \hline
		No. of UAVs (\emph{K}), UEs (\emph{U}), req. UEs (\emph{N})                              & 4 \cite{wang2023learning}, 100,                         1-30  \cite{gu2021association} \\ \hline
		Bandwidth (\emph{B})                                      & 20 MHz \cite{3gpp2018study}                       \\ \hline
		Channel bandwidth allocated to the                           & B/N MHz                                           \\
		requesting UEs (Bn)                                   &                                                   \\ \hline
		Max. trans. power of GBS, UAVs and RUEs               & 30, 27, 23 dBm  \\ \hline
		Antenna gain of the GBS, and UAVs                         & 3 dB                                              \\ \hline
		Noise spectral density                                    & -174 dBm/Hz                                       \\ \hline
		Mean interference from adjacent cells $I_{b} $            & -130 dBm/Hz  \cite{becvar2022energy}                                      \\ \hline
		Height of the GBS, UAVs, UEs antenna                      & 35, 100, 1.5 m                                    \\ \hline
		Zipf exponent $(\gamma)$                                      & 0.5  \cite{chen2021multi}                                              \\ \hline
		Content catalog of the GBS, and UAVs                      & 50, 10 files \cite{gao2023delay}                                      \\ \hline
		Content size                                              & 10 Mbits  \cite{zhang2023joint}                                          \\ \hline
		Number of simulation drops                                & 1000                                              \\ \hline
	\end{tabular}
\end{table}
\section{Simulation Results}

\begin{figure}[b!]
\centering
\includegraphics[scale=0.45]{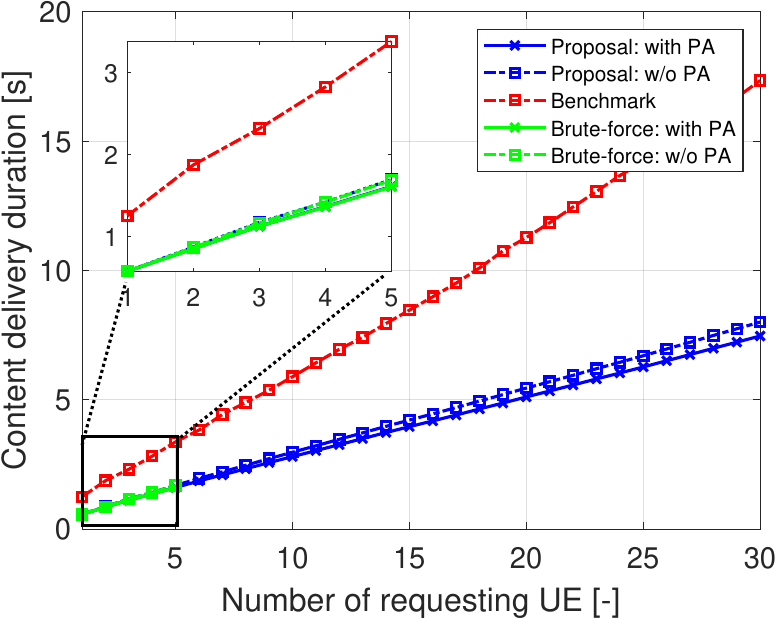}
\caption{Average content delivery over no. of req. UEs.} 
\label{fig:pdfresizer.com-pdf-crop (7)}
\end{figure}

Fig. \ref{fig:pdfresizer.com-pdf-crop (7)} illustrates the average content delivery durations achieved by different schemes for different number of requesting UEs. From Fig. \ref{fig:pdfresizer.com-pdf-crop (7)} can be observed that the content delivery duration of all schemes rises when the number of requesting UEs in the system increases. This is because bandwidth $B$ is divided among a larger number of requesting UEs. Still, the proposed algorithm always outperforms the benchmark algorithm. Take the 30 requesting UEs in Fig. \ref{fig:pdfresizer.com-pdf-crop (7)} as an example, compared to the benchmark algorithm, the proposed route selection algorithm yields 53.69\% gain in terms of reducing the average content delivery duration; this confirms the benefits of the proposed route selection algorithm. This is  because the benchmarking algorithm sends the contents sequentially, as shown in Fig. \ref{fig:Drawing2}a. Furthermore, employing the joint route selection and power allocation algorithm yields a 56.98\% gain over the benchmark algorithm. 

Fig. \ref{fig:pdfresizer.com-pdf-crop (7)} also demonstrates that the proposed algorithm provides close performance to the brute-force algorithm in terms of content delivery duration. Specifically, the gap between the proposed route selection algorithm without power allocation and the brute-force algorithm is only up to 0.52\%. Similarly, the gap between the proposed joint route selection and power allocation algorithm and the brute-force algorithm with power allocation is up to 0.30\%. The results presented herein provide evidence that the proposed greedy algorithm exhibits a strong correlation in route selection with the brute-force algorithm for content delivery.
\begin{figure}[t]
	\begin{minipage}[t]{0.5\linewidth}
		\includegraphics[width=\linewidth]{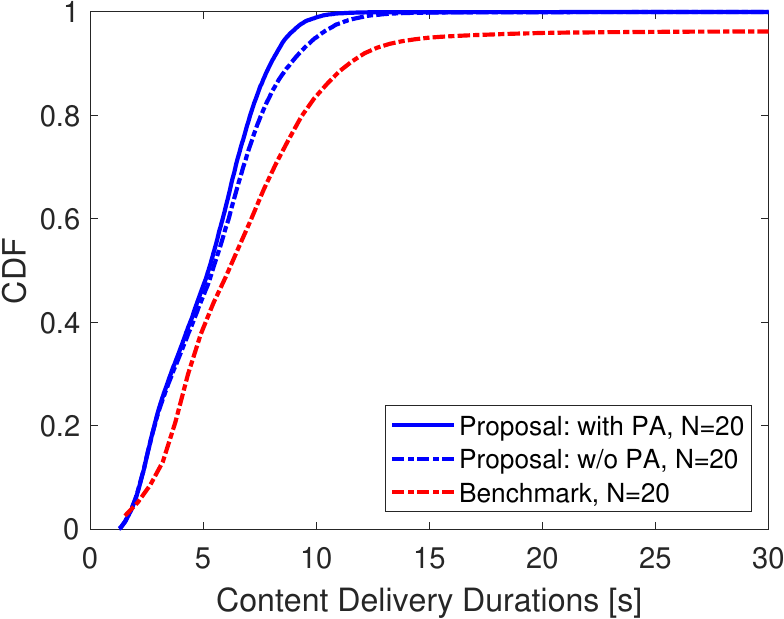}
		
		\label{pdfresizer.com-pdf-crop_UE_20}
	\end{minipage}%
	\hfill%
	\begin{minipage}[t]{0.5\linewidth}
		\includegraphics[width=\linewidth]{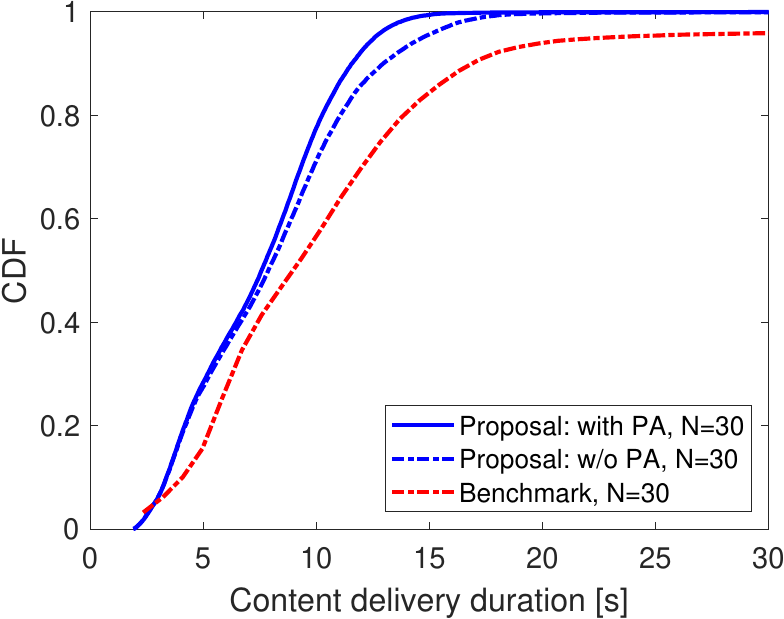}
		\label{pdfresizer.com-pdf-crop_UE_30}
	\end{minipage} 
	\caption{CDF of content delivery duration for 20 requesting UEs (left figure) and 30 requesting UEs (right figure).}
\end{figure}

While the previous figure showed the average content delivery duration, it does not give any insight on distribution of actual durations. Thus, we also the cumulative distribution function (CDF) of content delivery duration for individual algorithms in Fig. 4. Then, for any given confidence level, the confidence interval for the content delivery duration can be obtained from the CDF curve. Note that since we show results for 20 and 30 requesting UEs in Fig. 4a and Fig. 4b, respectively, we are not able to include brute-force algorithm due to its huge complexity. The benchmark algorithm's CDF curve in Fig. 4 exhibits a prolonged plateau that extends well beyond the content delivery durations of most UEs, so the CDF does not reaching 1 in given interval. This suggests that a small fraction of UEs experience significantly longer delivery durations compared to the majority, highlighting the benchmark algorithm's inefficiency in handling outlier cases. Fig. 4 further shows that the proposed route selection algorithm achieves lower content delivery duration than the benchmark algorithm. In particular, for a high number of requesting UEs, the proposed greedy algorithm produces more consistent results than the benchmark algorithm.  For instance, when N = 20 (Fig. 4a), roughly 99\% and 85\% of requested UEs, respectively, receive their content within 10 s using the proposed joint route selection and power allocation algorithm and the benchmark algorithm. When N = 30 (Fig. 4b), roughly 77\% and 55\% of requested UEs, respectively, receive their content within 10 s using the proposed joint route selection and power allocation algorithm and the benchmark algorithm. 

The result confirms that proposed scheme is an effective solution to defined problem. Besides, the more users are requesting the content, the more significant gap between proposed algorithm and benchmark. These are encouraging results considering the fact that, in general, number of UEs in the future 6G networks is assumed to grow exponentially.

\section{Conclusion}
In this article, we have shed light on the problem of joint route selection and power allocation in multi-hop- and cache-enabled networks. Since the defined problem is a mixed-integer non-linear programming problem, thus NP-hard, we have also designed a greedy-based algorithm efficiently managing jointly route selection and power allocation. We have demonstrated that the proposed greedy algorithm reduces overall content delivery duration by up to 56.98\% compared to the close related benchmark algorithm while it yields close-to-optimal performance. In the future, the proposed model can be extended by the optimization of power splitting among individual contents, incentivization of relaying users, or smart positioning of the unmanned aerial vehicles.

\bibliographystyle{IEEEtran}

\begin{thebibliography}{1}
\bibitem{zyrianoff2022iot} I. Zyrianoff, A. Trotta, L. Sciullo, F. Montori, and M. Di Felice, ``IoT edge caching: taxonomy, use cases and perspectives,'' \emph{IEEE Internet of Things Magazine}, vol. 5, no. 3, pp. 12-18, 2022.
	\bibitem{duong2022uav} T. Q. Duong, K. J. Kim, Z. Kaleem, M.-P. Bui, and N.-S. Vo, "UAV caching in 6G networks: A survey on models, techniques, and applications," \emph{Physical Communication}, vol. 51, p. 101532, 2022.
	\bibitem{zhao2019caching} N. Zhao \emph{et al.}, "Caching unmanned aerial vehicle-enabled small-cell networks: Employing energy-efficient methods that store and retrieve popular content," \emph{IEEE Veh. Technol. Mag.}, vol. 14, no. 1, 71-79, 2019.
	\bibitem{mach2022device} P. Mach, and Z. Becvar, "Device-to-device relaying: Optimization, performance perspectives, and open challenges towards 6G networks," \emph{IEEE Commun. Surv. Tutor.}, vol. 24, no. 3, pp. 1336-1393, 2022.
	\bibitem{ji2020joint} J. Ji \emph{et al.}, "Joint trajectory design and resource allocation for secure transmission in cache-enabled UAV-relaying networks with D2D communications," \emph{IEEE Internet Things J.}, vol. 8, no. 3, pp. 1557-1571, 2020.
	\bibitem{al2021throughput} M. S. Al-Abiad \emph{et al.}, "Throughput Maximization of Network-Coded and Multi-Level Cache-Enabled Heterogeneous Network,"  \emph{IEEE Trans. Veh. Technol.}, vol. 70, no. 10, pp. 11039-11043, 2021.
	\bibitem{qi2022joint} X. Qi, M. Yuan, Q. Zhang, and Z. Yang, "Joint power-trajectory-scheduling optimization in a mobile UAV-enabled network via alternating iteration,"  \emph{China Communications}, vol. 19, no. 1, pp. 136-152, 2022.
	\bibitem{zhang2018trajectory} G. Zhang, H. Yan, Y. Zeng, M. Cui, and Y. Lui, "Trajectory optimization and power allocation for multi-hop UAV relaying communications,"  \emph{IEEE Access}, vol. 6, pp. 48566-48576, 2018.
	\bibitem{wang2021deep} D. Wang \emph{et al.}, "Deep Reinforcement Learning for Caching in D2D-Enabled UAV-Relaying Networks,"  \emph{IEEE ICCC}, 2021, pp. 635-640.
	\bibitem{luo2023cost} L. Luo, R. Sun, R. Chai, and Q. Chen, "Cost-Efficient UAV Deployment and Content Placement for Cellular Systems With D2D Communications,"  \emph{IEEE Systems Journal}, 2023.
	\bibitem{zhang2023joint} T. Zhang, C. Chen, D. Yang, "Joint user association and caching placement for cache-enabling UAV networks,"  \emph{China Communications}, vol. 20, no. 6, pp. 291-309, 2023.
	\bibitem{blaszczyszyn2015optimal} B. Blaszczyszyn, and A. Giovanidis, "Optimal geographic caching in cellular networks,"  \emph{IEEE ICC}, 2015, pp. 3358-3363.
	\bibitem{gu2021association} Z. Gu \emph{et al.}, "Association and caching in relay-assisted mmWave networks: A stochastic geometry perspective,"  \emph{IEEE Trans. Wirel. Commun.}, vol. 20, no. 12, pp. 8316-8332, 2021.
	\bibitem{wang2023learning} J. Wang \emph{et al.}, "Learning-based dynamic connectivity maintenance for UAV-assisted D2D multicast communication,"  \emph{China Communications}, 2023.
	\bibitem{3gpp2018study}3GPP, "Study on channel model for frequencies from 0.5 to 100 GHz," 2018
		\bibitem{becvar2022energy} Z. Becvar, M. Nikooroo, P. Mach, "On energy consumption of airship-based flying base stations serving mobile users,"  \emph{IEEE Trans. Commun.}, vol. 70, no. 10, pp. 7006-7022, 2022.
	\bibitem{chen2021multi} Y.-J. Chen \emph{et al.}, "Multi-agent reinforcement learning based 3D trajectory design in aerial-terrestrial wireless caching networks,"  \emph{IEEE Trans. Veh. Technol.}, vol. 70, no. 8, pp. 8201-8215, 2021.
	\bibitem{gao2023delay} X. Gao, Z. Qian, X. Wang, "Delay-Oriented Probabilistic Edge Caching Strategy in a Device-to-Device Enabled IoT System,"  \emph{IEEE Sens. J.}, 2023.
	\bibitem{mach2021power} P. Mach, Z. Becvar, and M. Najla, "Power allocation, channel reuse, and positioning of flying base stations with realistic backhaul,"  \emph{IEEE Internet Things J.}, vol. 9, no. 3, pp. 1790-1805, 2021.
\end{thebibliography}

\end{document}